\documentclass[12pt,preprint]{aastex}
\usepackage{epsfig}
\usepackage{natbib}
\usepackage{graphicx}
\usepackage{slashbox}
\usepackage{multirow}
\usepackage{lscape}
\usepackage{mathrsfs,amssymb}
\usepackage{amsmath}
\usepackage{subfigure}
\usepackage{amssymb}
\usepackage{multirow}
\usepackage{tabularx}
\usepackage{rotating}
\usepackage{amsmath}
\usepackage{ulem}
\usepackage{lineno}
\usepackage{booktabs}  

\newcommand       \Angstrom     {\,{\rm \AA}}

\newcommand       \cm           {\,{\rm cm}}

\newcommand       \eV           {\,{\rm eV}}
\newcommand       \keV           {\,{\rm keV}}
\newcommand       \g            {\,{\rm g}}

\newcommand       \NH           {N_{\rm H}}
\newcommand       \nH           {n_{\rm H}}
\newcommand       \simlt        {\lesssim}

\newcommand       \gtsim        {\gtrsim}

\newcommand       \mum          {\,{\rm \mu m}}
\newcommand       \ppm          {\,{\rm ppm}}

\newcommand       \Msun         {\,{M_\odot}}

\newcommand       \simali       {\sim\,}
\newcommand       \magni        {\,{\rm mag}}
\newcommand       \rmH          {{\rm H}}
\newcommand       \rmHI          {{\rm HI}}
\newcommand       \rmHH        {{\rm H_2}}
\newcommand{\bigsigma}{\mbox{\Large \ensuremath{\sigma}}}
\def	\sigmadustabs {{{\bigsigma}_{\rm dust}^{\rm abs}}}
\def	\sigmadustsca {{{\bigsigma}_{\rm dust}^{\rm sca}}}
\def	\sigmadust {{{\bigsigma}_{\rm dust}^{\rm ext}}}
\def	\sigmagas {{{\bigsigma}_{\rm gas}^{\rm abs}}}
\def	\amin	  {a_{\rm min}}
\def	\amax        {a_{\rm max}}
\def  \xism    {{\rm\left(\rm X/H\right)}_{\rm ISM}}
\def    \siism {{\rm\left(\rm Si/H\right)}_{\rm ISM}}
\def    \mgism {{\rm\left(\rm Mg/H\right)}_{\rm ISM}}
\def    \feism {{\rm\left(\rm Fe/H\right)}_{\rm ISM}}
\def    \cism   {{\rm\left(\rm C/H\right)}_{\rm ISM}}

\def    \xgas   {{\rm\left(\rm X/H\right)}_{\rm gas}}

\def    \cgas    {{\rm\left(\rm C/H\right)}_{\rm gas}}
\def    \xdust   {{\rm\left(\rm X/H\right)}_{\rm dust}}
\def    \mgdust   {{\rm\left(\rm Mg/H\right)}_{\rm dust}}
\def    \fedust   {{\rm\left(\rm Fe/H\right)}_{\rm dust}}
\def    \sidust   {{\rm\left(\rm Si/H\right)}_{\rm dust}}
\def    \cdust    {{\rm\left(\rm C/H\right)}_{\rm dust}}
\def    \odust    {{\rm\left(\rm O/H\right)}_{\rm dust}}

\def    \Vs                {V_{\rm s}}
\def    \Vsmax         {V_{\rm s,max}}
\def    \Vg                {V_{\rm g}}
\def    \Vgmax         {V_{\rm g,max}}
\newcommand{\avg}[1]{\ensuremath{\langle #1 \rangle}}
\newcommand{\Nbar}{\ensuremath{\avg{N}}}
\newcommand{\sigbar}{\ensuremath{\avg{\sigma}}}
\newcommand{\dd}{\ensuremath{{\rm d}}}
\newcommand       \be           {\begin{equation}}
\newcommand       \ee           {\end{equation}}
\shorttitle{Interstellar X-ray Absorption and Scattering}
\title{
\vspace*{-2.0em}
{\normalsize\rm Accepted for publication in
               {\it The Astrophysical Journal}}\\
\vspace*{1.0em}
Interstellar X-ray Absorption and Scattering
\\{\small DRAFT: \today ~~}
}
\author{Linli~Yan\altaffilmark{1,2,3},
             Aigen Li\altaffilmark{2}
             and Fangjun~Lu\altaffilmark{3}
             }
\altaffiltext{1}{School of Mathematics and Physics,
                       Anhui Jianzhu University, 
                       Hefei, Anhui 230601, China;
                       {\sf yan.linli@foxmail.com}}
\altaffiltext{2}{Department of Physics and Astronomy,
                        University of Missouri,
                        Columbia, MO 65211, USA;
                        {\sf lia@missouri.edu}}
\altaffiltext{3}{Key Laboratory of Particle Astrophysics,
                       Institute of High Energy Physics,
                       Chinese Academy of Sciences,
                       Beijing 100049, China;
                       {\sf lufj@ihep.ac.cn}}            

\begin{document}

\begin{abstract}
Accurate estimates of the absorption of X-rays
by interstellar gas and dust are of crucial importance
for the analysis and interpretation of almost all
astronomical soft X-ray observations.
However, the present X-ray absorption data
extensively used by the community were derived
from a reduced interstellar abundance
($\simali$70\% of solar) and ignoring dust scattering. 
Therefore, these X-ray absorption data,
although highly popular, could have been
substantially underestimated. Here we update
the interstellar X-ray absorption and scattering
by making use of updated atomic cross sections,
updated interstellar abundances,
and realistic X-ray dust physics,
and appropriately distributing 
metal elements in gas and dust.
The resulting X-ray absorption and scattering
data are publicly available on {\tt GitHub}.
\end{abstract}


\section{Introduction}\label{sec:intro}
It is well recognized that the absorption (and scattering)
of X-rays by interstellar material alters the X-ray spectra
of almost all cosmic X-ray sources. In order to interpret
the observed X-ray spectra, an accurate knowledge of
the interstellar X-ray absorption (and scattering)
is of crucial importance for correcting for the alteration
caused by the intervening interstellar gas and dust.

Various studies have been carried out to estimate
the X-ray absorption of the interstellar medium (ISM)
since the pioneering work of Strom \& Strom (1961)
who considered the absorption of {\it gaseous}, atomic 
H, He and other thirteen elements, including C, N, O,
Ne, Na, Mg, Al, Si, S, Ar, Ca, Fe and Ni.
Strom \& Strom (1961) adopted 
the solar abundances of Cameron (1959) for these
elements and assumed the atomic absorption coefficients
to be a simple power-law of wavelength. 
Subsequent improvements have been made by 
Felten \& Gould (1966), 
Bell \& Kingston (1967),
Brown \& Gould (1970), 
Morrison \& McCammon (1983), and 
Ba\l{}uci\'{n}ska-Church \& McCammon (1992)
who considered a larger number of 
astrophysically important elements 
and used the then more up-to-date solar abundances 
and more accurate atomic absorption cross sections.  

Fireman (1974) was the first to examine the effects of 
{\it solid} interstellar dust on X-ray absorption.
He argued that the X-ray absorption would be reduced
if a large fraction of the heavy atoms 
in the ISM are bound in dust because of
the ``self-shielding'' effects of dust:
as the X-ray optical depth of a large grain
can be much greater than unity,
most of the X-ray absorption
will occur on its surface. 
In other words,  the atoms in the interior 
of a dust grain are ``blanketed'' and 
would reduce the effective absorptivity 
of the ISM relative to 
that of a completely gaseous medium.
Ride \& Walker (1977) further explored 
the ``self-shielding'' effects 
and computed the X-ray absorption of the ISM
in two phases: a cool, dense cloud phase 
and a hot, tenuous intercloud phase.

With the advent of the Chandra X-ray Observatory
in 1999, an accurate knowledge of the interstellar
X-ray absorption was urgently needed.
To this end, efforts have been made by
Wilms et al.\ (2000) who computed the X-ray
absorption cross sections of interstellar dust and gas. 
They used the (then) updated 
photoionization cross section of atoms and H$_2$.
Unlike previous studies in which the ISM was assumed
to have a solar chemical composition, 
Wilms et al.\ (2000) adopted a ``reduced'' 
or ``subsolar'' abundance for the ISM: they 
assumed that the interstellar abundances
are only $\simali$70\% of solar.

The X-ray interstellar absorption cross sections of 
Morrison \& McCammon (1983),  
Ba\l{}uci\'{n}ska-Church \& McCammon (1992), and
Wilms et al.\ (2000) have been used extensively 
by the X-ray community. Their results were
incorporated in the {\tt XSPEC} X-ray fitting 
package (Arnaud 1996) as 
{\tt Wabs} (Morrison \& McCammon 1983),
{\tt Phabs} (Ba\l{}uci\'{n}ska-Church \& McCammon 1992),
and {\tt TBabs} (Wilms et al.\ 2000), respectively.
However, as mentioned earlier,
Morrison \& McCammon (1983) 
and Ba\l{}uci\'{n}ska-Church \& McCammon 1992
neglected contributions from ions, molecules and dust.
Subsequently, Gatuzz et al.\ (2015) presented 
an X-ray absorption model for the ISM.
They considered both neutral and ionized gas species,
but neglected dust.
Presently, to our knowledge, the standard interstellar
X-ray absorption data extensively used by the X-ray
community were that of Wilms et al.\ (2000),
computed from {\tt TBabs} 
(and its new version {\tt TBnew}).

It has been 25 years since Wilms et al.\ (2000) 
published their X-ray absorption data.
There has been considerable progress in 
both experimental measurements and theoretical
calculations of the photoelectric absorption 
properties of atoms, ions and solids
(e.g., see Gatuzz et al.\ 2015).
There is also an improved understanding of
the interstellar abundances and dust models.  
Therefore, it is time to make use of 
these advances to update the X-ray absorption 
of interstellar dust and gas.
In particular, we argue that Wilms et al.\ (2000)
had considerably underestimated the interstellar
X-ray absorption. As elaborated below, the true
interstellar abundance cannot be subsolar
(see \S\ref{sec:abundances}). Furthermore,
in addition to absorption, solid dust grains
also {\it scatter} X-ray photons while Wilms et al.\ (2000)
ignored scattering (see \S\ref{sec:scattering}).

\subsection{What are the true interstellar elemental abundances?}\label{sec:abundances}
What might be the most appropriate set of interstellar 
abundances of metal elements (both in gas and in dust)
relative to hydrogen has been a subject of much discussion
in the past decades (see Snow \& Witt 1996, Sofia 2004,
Li 2005, Jenkins 2009, Wang et al.\ 2015, Zuo et al.\ 2021). 
Historically, the interstellar abundances of
metal elements like C, O, Mg, Si, and Fe were 
commonly assumed to be solar. 
In the late 1990s, it was argued that,
because of their young ages,
the interstellar abundances might be better
represented by those of B stars and young F, G stars, 
which are just $\simali$60--70\%  
of the solar values 
(Snow \& Witt 1996, Sofia \& Meyer 2001).
This led Wilms et al.\ (2000) to
assume a ``subsolar'' abundance.
However, as demonstrated by Li (2005)
and more recently by Zuo et al.\ (2021),
if the interstellar abundances are indeed
``subsolar'' like those adopted by Wilms et al.\ (2000), 
there will be a shortage of raw material to 
form the dust to account for the observed 
ultraviolet (UV), optical, and infrared (IR)
interstellar extinction.

We also note that the abundances of B stars,
on which the whole ``subsolar'' idea was based,
have also undergone appreciable changes.
Przybilla et al.\ (2008) 
and Nieva \& Przybilla (2012) derived
the photospheric abundances of heavy elements
for unevolved early B-type stars
using the more realistic non-local thermodynamic
equilibrium (NLTE) techniques.
As shown in Table~\ref{tab:abund},   
they found that the photospheric abundances of 
those B stars are in close agreement 
with the updated, widely-used solar abundances
of Asplund et al.\ (2009).

It is also worth noting that
the solar abundances reported
in the literature have also undergone
major changes over the years
(see Table~\ref{tab:abund}).
The solar abundances compiled 
by Asplund et al.\ (2009)  
were significantly reduced
from their earlier values
(e.g., Anders \& Grevesse 1989).
More recently, Asplund et al.\ (2021)
re-assessed the solar elemental abundances
and found that Si is lower by $\simali$30\%
than that of Asplund et al.\ (2009),
while other elements such as C, O,
Mg, and Fe are consistent with that
of Asplund et al.\ (2009) within a few
percent (see Table~\ref{tab:abund}).
In contrast, Lodders et al.\ (2025) derived
much higher abundances for C and O
which exceed that of Asplund et al.\ (2009)
by $\simali$20\%, while the abundances of
Mg, Si, and Fe generally agree with that of
Asplund et al.\ (2009).

Furthermore, Lodders (2003) argued that
the currently observed solar photospheric
abundances must be lower than those of
the proto-Sun because helium and other
heavy elements have settled toward the Sun's
interior since the time of the Sun's formation 
$\simali$4.55\,Gyr ago.
The latest proto-Sun abundances
of C, O, Mg, Si, and Fe determined by
Lodders et al.\ (2025) are considerably
higher than that of Asplund et al.\ (2009),
by $\simali$48\%, 45\%, 15\%, 35\%,
and 20\%, respectively
(see Table~\ref{tab:abund}).

Also, over the past 4.55\,Gyr, the ISM has been
enriched with metals, primarily through the life
and death of stars.
The Galactic chemical enrichment (GCE)
could have led the abundances of C, O, Mg,
Si and Fe to an increase of $\simali$0.06,
0.04, 0.04, 0.08 and 0.14 dex, respectively
(see Chiappini et al.\ 2003).
We take the GCE-augmented protosolar
abundances of Lodders et al.\ (2025)
to represent the interstellar abundances 
in the solar neighborhood.
As shown in Table~\ref{tab:abund},
the reduced abundances of Wilms et al.\ (2000)
are substantially lower than that of solar, 
proto-Sun, proto-Sun\,+\,GCE, and even B stars.
More specifically, the proto-Sun\,+\,GCE
abundances of C, O, Mg, Si, and Fe
which we adopt as the interstellar abundances
considerably exceed those adopted by
Wilms et al.\ (2000) by $\simali$90\%, 58\%,
100\%, 182\%, and 95\%, respectively.
Therefore, Wilms et al.\ (2000) must have
underestimated the X-ray absorption.

\begin{sidewaystable}
\centering
\footnotesize
\caption{\label{tab:abund}
Solar and stellar abundances for
the major dust-forming elements
(relative to $10^6$ H atoms).}
\begin{tabular}{lcccccccccc}
\hline \hline
Element  & Wilms et al.\ (2000)
         & B stars$^{a}$
         & Sun$^{b}$ 
         & Sun$^{c}$
         & Sun$^{d}$
         & Sun$^{e}$
         & Proto-Sun$^{f}$
         & Proto-Sun$^{e}$
         & Proto-Sun\,+\,GCE$^{g}$\\
\hline
C  & 240
   & $209\pm15$
   & $363\pm33$
   & $269\pm31$
   & $288\pm27$
   & $324\pm67$
   & $288\pm27$
   & $398\pm83$
   & $457\pm95$\\
O  & 490
   & $575\pm40$
   & $851\pm69$
   & $490\pm57$
   & $490\pm45$
   & $575\pm66$
   & $575\pm66$
   & $708\pm82$
   & $776\pm89$\\   
Mg & 25.1
   & $36.3\pm4.2$
   & $38.0\pm4.4$
   & $39.8\pm3.7$
   & $35.5\pm2.5$
   & $37.2\pm0.9$
   & $41.7\pm1.9$
   & $45.7\pm1.1$
   & $50.1\pm1.2$\\
Si & 18.6
   & $31.6\pm1.5$
   & $35.5\pm4.1$
   & $32.4\pm2.2$
   & $22.4\pm2.2$
   & $35.5\pm0.8$
   & $40.7\pm1.9$
   & $43.7\pm1.0$
   & $52.5\pm1.2$\\
Fe & 26.9
   & $27.5\pm2.5$
   & $46.8\pm3.2$
   & $31.6\pm2.9$
   & $28.8\pm2.7$
   & $30.9\pm0.7$
   & $34.7\pm2.4$
   & $38.0\pm0.9$
   & $52.5\pm1.2$\\
\hline
\end{tabular}
(a) Przybilla et al.\ (2008). 
(b) Anders \& Grevesse (1989). 
(c) Asplund et al.\ (2009). 
(d) Asplund et al.\ (2021).  
(e) Lodders et al. \ (2025).   
(f) Lodders (2003).
(g) GCE-augmented protosolar
abundances of Lodders et al.\ (2025),
with an enrichment of $\simali$0.06,
0.04, 0.04, 0.08, and 0.14 dex
for C, O, Mg, Si, and Fe, respectively
(Chiappini et al.\ 2003).
\end{sidewaystable}

\subsection{Dust as an X-ray scatterer}\label{sec:scattering}
Wilms et al.\ (2000) used the ``self-shielding'' 
approximation to estimate the X-ray absorption 
cross section of dust per H nucleon:
\begin{equation}\label{eq:sigmadust}
\bigsigma_{\rm dust}/\rmH = \xi_{\rm gr}  
\int_0^\infty \frac{\dd n}{\dd a}\,\pi a^2
\left\{1-\exp\left[-\sigbar \Nbar\right]\right\}\,\dd a ~~,
\end{equation}
where $\xi_{\rm gr}$ is the number of grains
per H atom along the line of sight,
$\dd n/\dd a$ is the dust size distribution,  
$\sigbar$ is the average photoionization 
cross section of the dust material, and
$\Nbar$ is the average column densities of atoms 
(measured in atoms per cm$^2$) of the dust material.
This approximation assumes that the self-blanketing 
is the same for each element, and more importantly,
it ignores the scattering of X-rays by dust.
As dust is not only an X-ray absorber
but also an X-ray scatterer, Wilms et al.\ (2000) would
therefore substantially underestimate the X-ray attenuation.

In this work, we aim to calculate the X-ray
absorption and scattering of interstellar dust
and gas as a function of energy
in the 0.1--10$\keV$ range, using 
(1) the most accurate atomic data available,
(2) the most recent abundance determinations,
and (3) realistic X-ray scattering and absorption
properties of dust.
This paper is organized as follows.
We first formulate the interstellar X-ray scattering
and absorption in \S\ref{sec:formulating}.
As the X-ray scattering and absorption of
dust grains depend on their sizes, we derive
the dust size distributions in \S\ref{sec:dust}
by fitting the Galactic average interstellar extinction curve.
\S\ref{sec:gas} discusses the X-ray absorption
arising from gaseous species.
The resulting interstellar X-ray scattering
and absorption are presented and discussed
in \S\ref{sec:results}.
Finally, the major results are summarized
in \S\ref{sec:summary}.

\section{Formulating the Interstellar
           X-ray Absorption}\label{sec:formulating}
When the X-rays emitted by an X-ray source
pass through the ISM, because of the absorption
and scattering by the interstellar gas and dust,
the intensities of the X-rays will be reduced
and their spectra will be altered.
Let $I_{\rm obs}(E)$ and $I_\bigstar(E)$
be the observed X-ray intensity
(as a function of energy $E$)
and that emitted by an X-ray source, respectively.
Apparently, $I_{\rm obs}(E)$ and $I_\bigstar(E)$
are related through
\begin{equation}
I_{\rm obs}(E) = I_\bigstar(E) 
\exp\left\{-\NH\left(\sigmadust/\rmH + 
\sigmagas/\rmH\right)\right\} ~~,
\end{equation}
where $\NH$ is the hydrogen column density
of the ISM, $\sigmagas/\rmH$ is the gas absorption 
cross section per H nucleon, and
$\sigmadust/\rmH$ is the dust {\it extinction}
cross section per H nucleon.

The dust extinction cross section $\sigmadust$
is a combination of absorption and scattering.
It depends on the dust composition, size distribution
and morphological structure.
We will assume the interstellar dust to be a mixture
of two grain types---compact, spherical amorphous
silicate grains and carbonaceous grains---each with
an extended size distribution.
Following Li \& Draine (2001), we assume
that the carbonaceous grain population extends from
grains with graphitic properties at radii $a\gtsim0.01\mum$,
down to particles with polycyclic aromatic hydrocarbon
(PAH)-like properties at very small sizes.
The quantities and size distributions of both dust types
will be determined by fitting the interstellar extinction
curve (see \S\ref{sec:dust}). This also determines
$\xdust$, the amount of element X (per H nucleon)
required to be locked up in dust,
where X represents C, O, Si, Mg, and Fe,
the major dust-forming elements.

The gas absorption cross section (per H nucleon),
$\sigmagas/\rmH$, is obtained by summing
the photoionization cross sections
of individual atoms, ions, and molecules,\footnote{%
  \label{ftnt:fracH2}
  For molecules, following Wilms et al.\ (2000),
  we only consider molecular hydrogen
  because of its large abundance.
  In the Milky Way galaxy,
  the total atomic hydrogen mass
  is $\simali$$2.9\times10^9\Msun$
  and the total molecular hydrogen mass
  is $\simali$$8.4\times10^8\Msun$
  (see Draine 2011).
  If we ignore ionized hydrogen,
  we estimate an atomic hydrogen
  fraction (by number) of $\simali$90\%
  and a molecular hydrogen
  fraction of $\simali$10\%.
  }
weighting their contributions by their abundances
(see \S\ref{sec:gas}):
\begin{equation}\label{eq:xray_gas_abs_1}
\sigmagas/\rmH = \sum_{{\rm X}}\sum_{i}
\xism\left(1-\beta_{\rm X}\right)
{a_{{\rm X},i}}
{\bigsigma_{\rm bf}({\rm X},i)}
\,+\,f({\rmHI})\bigsigma_{\rm bf}(\rmHI)
\,+\,f(\rmHH)\bigsigma_{\rm bf}(\rmHH) ~~,
\end{equation} 
where $\xism$ is the total interstellar abundance
of element X (relative to H) that is heavier than H,
$\beta_{\rm X}\equiv \xdust/\xism$ is the ratio
of the abundance of element X in dust to its total
interstellar abundance,\footnote{%
   The gas-phase abundance of element X
   is obtained from $\xgas=\xism-\xdust$.
   Then, $\left(1-\beta_{\rm X}\right) = \xgas/\xism$
   is the so-called ``depletion factor''
   as defined by Wilms et al.\ (2000).
   }
$a_{{\rm X},i}$ is the fraction of ions of
element X that are in ionization stage $i$,
$\bigsigma_{\rm bf}({\rm X},i)$ is the total photoionization
cross section of element X in ionization stage $i$,
$f(\rmHI)\approx0.9$ and $f(\rmHH)\approx0.1$
are respectively the relative amounts of atomic hydrogen
($\rmHI$) and H$_{2}$ with respect to the total hydrogen
nucleons (see Footnote~\ref{ftnt:fracH2}),
$\bigsigma_{\rm bf}(\rmHI)$ is the photoionization
cross section of $\rmHI$,
and $\bigsigma_{\rm bf}({\rm H_2})$ is photoionization
cross section of H$_2$.

\section{Dust as an X-ray Absorber and Scatterer}\label{sec:dust}           
As mentioned in \S\ref{sec:formulating},
we assume chemically homogeneous
compact spherical grains
consisting of amorphous silicates, graphite, 
and polycyclic aromatic hydrocarbons.
Their contributions to the X-ray absorption
($\sigmadustabs$), scattering ($\sigmadustsca$),
and extinction ($\sigmadust$) at energy $E$,
on a per H nucleon basis, are
\begin{equation}\label{eq:xray_dust_abs}
\sigmadustabs/\rmH 
= \sum_{i} \int_{\amin}^{\amax} C_{{\rm abs},i}(a,E)
\frac{1}{\nH}\frac{\dd n_i}{\dd a}\dd a ~~,
\end{equation}
\begin{equation}\label{eq:xray_dust_sca}
\sigmadustsca/\rmH 
= \sum_{i} \int_{\amin}^{\amax} C_{{\rm sca},i}(a,E)
\frac{1}{\nH}\frac{\dd n_i}{\dd a}\dd a ~~,
\end{equation}
\begin{equation}\label{eq:xray_dust_ext}
\sigmadust/\rmH = \sigmadustabs/\rmH
+ \sigmadustsca/\rmH ~~,
\end{equation}
where the summation is over two different
grain materials
(i.e., $i=1$ refers to amorphous silicate dust
and $i=2$ refers to graphite and PAHs),
$\amin=3.5\Angstrom$ and $\amax=5\mum$
are the lower and upper cutoff grain sizes, respectively,
$C_{{\rm abs},i}(a,E)$ and $C_{{\rm sca},i}(a,E)$
are the absorption and scattering cross sections
at energy $E$ of a grain of size $a$ composed of
the $i$-th grain type,
$\dd n_i$ is the number density of grains
of the $i$-th type in the size range of $a$ and $a+\dd a$,
and $n_{\rm H}$ is the number density of H nuclei
(in both atoms and molecules).

We determine the size distributions ($dn/da$)
of both grain types from fitting the interstellar
extinction curve from the near-IR to the far-UV. 
We adopt the functional forms of Weingartner
\& Draine (2001; hereafter WD01):
\be
\frac{1}{\nH}
\left(\frac{\dd n}{da}\right)_{\rm g}
= D(a) + \frac{C_{\rm g}}{a} 
\left( \frac{a}{a_{\rm t,g}} \right)^{\alpha_{\rm g}}
F(a; \beta_{\rm g}, a_{\rm t,g}) \times
\begin{cases}
1~, & 3.5 \Angstrom < a < a_{\rm t,g}
\cr
\exp \left\{ - [(a - a_{\rm t,g})/a_{\rm c,g}]^3 \right\}~,
  & a > a_{\rm t,g}
\end{cases}
\label{eq:dnda_gra}
\ee
for carbonaceous dust (i.e., graphite and PAHs), and
\be
\frac{1}{\nH}
\left(\frac{\dd n}{da}\right)_{\rm s}
= \frac{C_{\rm s}}{a}
\left( \frac{a}{a_{\rm t,s}} \right)^{\alpha_{\rm s}} 
F(a; \beta_{\rm s}, a_{\rm t,s}) \times
\begin{cases}
1~, &3.5 \Angstrom < a < a_{\rm t,s}
\cr
\exp \left\{ - [(a - a_{\rm t,s})/a_{\rm c,s}]^3 \right\}~,
&a > a_{\rm t,s}
\end{cases}
\label{eq:dnda_sil}
\ee
for silicate dust.  
The $F(a; \beta, a_{\rm t})$ term is defined as
\be
F(a; \beta, a_{\rm t}) \equiv
\begin{cases}
1+\beta\,a/a_{\rm t}~~~, &\beta \ge 0
\cr
\left(1-\beta\,a/a_{\rm t} \right)^{-1}~~~, &\beta < 0 ~~.
\end{cases}
\ee
The $D(a)$ term is for the very small carbonaceous grains
(i.e., PAHs) and consists of two log-normal distribution
functions:
\be
\label{eq:lognormal}
D(a) = \sum_{i=1}^2
\frac{B_i}{a} 
\exp \left\{ - \frac{1}{2} \left[ \frac{\ln (a/a_{0,i})}
{\sigma_i} \right]^2 \right\}
~~~,~~~~~~~a > 3.5 \Angstrom 
\ee
\be
\label{eq:B}
B_i = \frac{3}{(2\pi)^{3/2}}
\frac{\exp(-4.5\sigma_i^2)}{\rho a_{0,i}^3\sigma_i}
\frac{
        b_{{\rm C},i}m_{\rm C} 
        }
        {
        1 + {\rm erf}[3 \sigma_i / \sqrt{2} + 
        \ln (a_{0,i} / 3.5 \Angstrom) / \sigma_i \sqrt{2}]
        }~~~,
\ee
where
$m_{\rm C}$ is the mass of a C atom,
$\rho=2.24\g\cm^{-3}$ is the density of graphite,
$b_{{\rm C},1}=45\ppm$ and $b_{{\rm C},2}=15\ppm$
are the C abundance (per H nucleus) in the two
log-normal populations,
$a_{0,1} = 3.5\Angstrom$,
$\sigma_1 = 0.40$,
$a_{0,2} = 20\Angstrom$,
and $\sigma_2 = 0.55$.\footnote{%
  \citet{WD01} took $a_{0,1} = 3.5\Angstrom$,
    $a_{0,2} = 30\Angstrom$,
    and $\sigma_1 = \sigma_2 =0.40$.
    These parameters were determined
    by Li \& Draine (2001) from modeling
    the near- and mid-IR emission of
    the Galactic diffuse ISM.
    Later, as the optical properties of PAHs, 
    graphite, and amorphous silicates
    adopted by Li \& Draine (2001) 
    were somewhat modified to be consistent
    with subsequent experimental measurements
    and {\it Spitzer} observations,
    Draine \& Li (2007) found that
    $a_{0,1} = 3.5\Angstrom$,
    $\sigma_1 = 0.40$,
    $a_{0,2} = 20\Angstrom$,
    and $\sigma_2 =0.55$
    are better at explaining the observed
    near- and mid-IR emission. 
    }

With the parameters for the two log-normal populations
pre-selected, we now have a total of 10 adjustable parameters:
$C_{\rm g}$, $a_{\rm t,g}$, $a_{\rm c,g}$, $\alpha_{\rm g}$,
and $\beta_{\rm g}$ for the carbonaceous component,
and $C_{\rm s}$, $a_{\rm t,s}$, $a_{\rm c,s}$, $\alpha_{\rm s}$,
and $\beta_{\rm s}$ for the silicate component.

\begin{figure*}[htp]
\vspace{3mm}
\begin{center}
\includegraphics[width=10cm,angle=0]{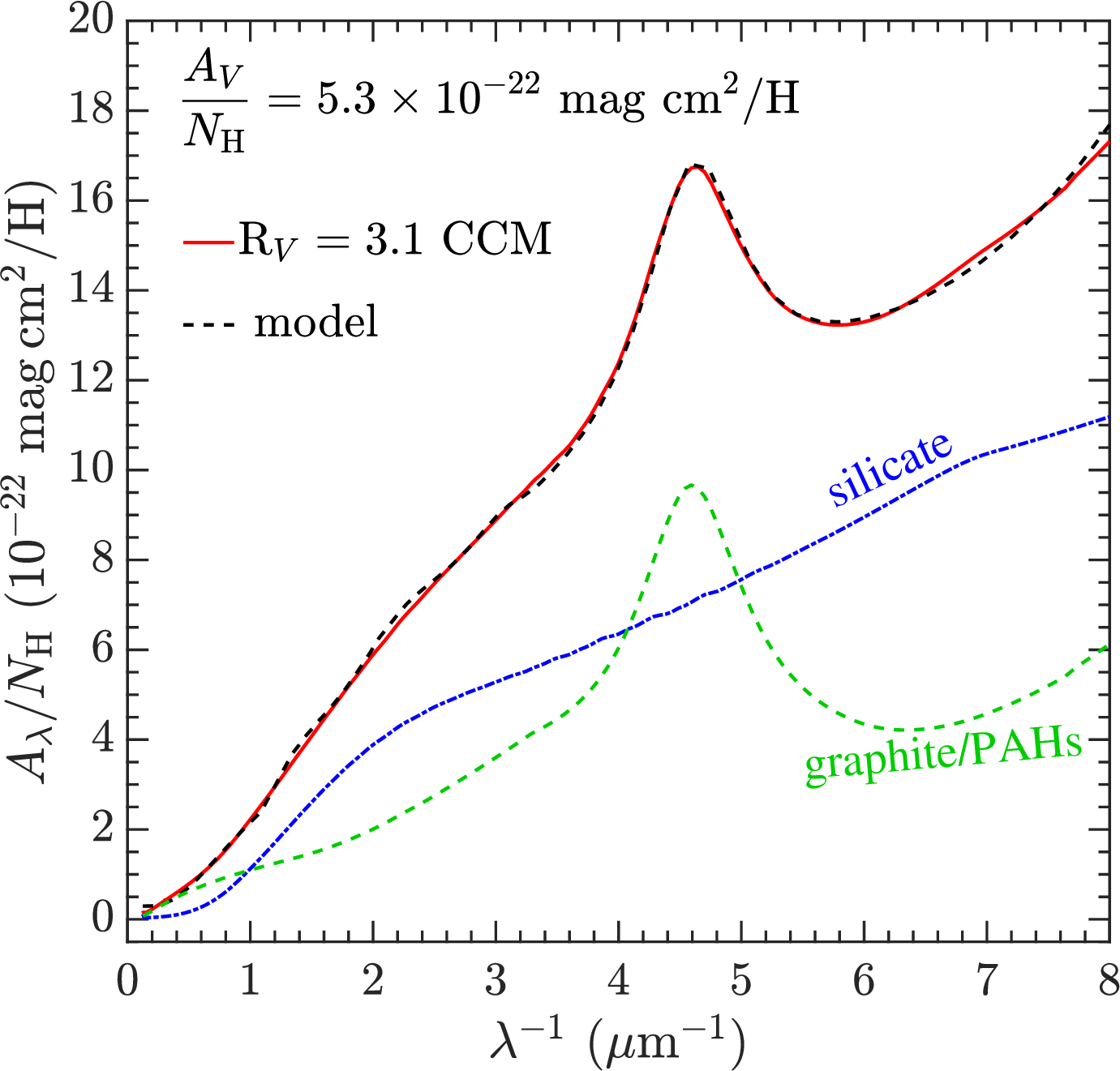}
\end{center}
\vspace{-4mm}
\caption{\label{fig:bestfit}
Fitting the Galactic average extinction curve
represented by the CCM $R_V=3.1$ curve
(solid red line) with a mixture of silicate
(blue dotted line) and graphite/PAHs
(green dashed line), each with an extended size
distribution (see Figure~\ref{fig:dnda}).
The model extinction curve---the sum of the silicate
and graphite/PAHs---is shown as a black dashed line.
         }
\end{figure*}

We fit the Galactic average interstellar extinction curve
in the wavelength range of 0.125 to 8$\mum$,
approximated by the parametrization of Cardelli, Clayton,
\& Mathis (1989; hereafter CCM) with $R_V=3.1$,
where $R_V\equiv A_V/E(B-V)$
is the optical total-to-selective extinction ratio,
$E(B-V)\equiv A_B-A_V$ is the reddening,  $A_V$ and
$A_B$ are the $V$- and $B$-band extinction,respectively.
We divide this wavelength range into 120 different
wavelengths, equally spaced in a logarithm scale.
The extinction is calculated in a manner similar to that
for X-ray absorption and scattering
(see eqs.\,\ref{eq:xray_dust_abs}--\ref{eq:xray_dust_ext}): 
\begin{equation}\label{eq:extcurv}
A_\lambda/\NH = 1.086 \sum_i
\int_{\amin}^{\amax} C_{{\rm ext},i}(a,\lambda)
\frac{1}{\nH}\frac{\dd n_i}{\dd a}\dd a~~,
\end{equation}
where the summation is, again, over the two grain types
(i.e., amorphous silicate and graphite/PAHs),
$A_\lambda$ is the extinction at wavelength $\lambda$,
$\NH$ is the hydrogen column density,
and $C_{{\rm ext},i}(a,\lambda)$
is the extinction cross section of
grain type $i$ of size $a$
at wavelength $\lambda$.
For the Galactic diffuse ISM,
we take the mean extinction-to-gas ratio of
$A_V/\NH=5.3\times10^{-22}\magni\cm^2\,\rmH^{-1}$
(Bohlin et al.\ 1978).
%

\begin{figure*}[htp]
\vspace{3mm}
\begin{center}
\includegraphics[width=10cm,angle=0]{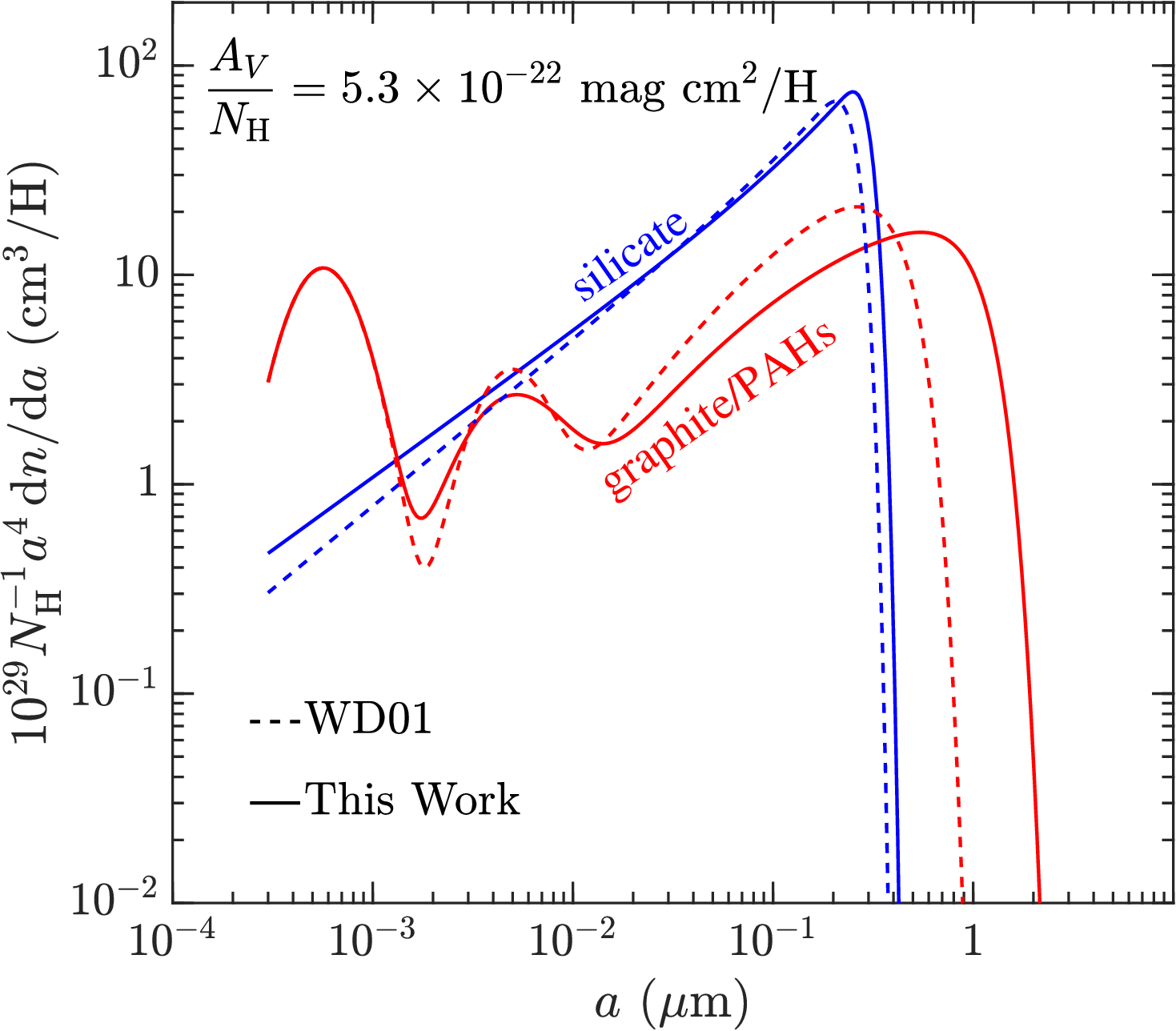}
\end{center}
\vspace{-4mm}
\caption{\label{fig:dnda}
Size distributions of amorphous silicates
(blue solid line) and graphite/PAHs (red dashed line)
derived from fitting the Galactic average extinction curve
of CCM $R_V=3.1$ (see Figure~\ref{fig:bestfit}).
For comparison, we also show as dashed lines
the original size distributions
for CCM $R_V=3.1$ derived by WD01.}
\end{figure*}

We calculate the extinction cross section
using Mie theory (Bohren \& Huffman 1983)
and the optical properties of ``astronomical''
silicates and carbonaceous materials
(i.e., graphite and PAHs) from \citet{Draine2007}. 
In fitting the near-IR to far-UV extinction curve,
we follow the approach of WD01 by measuring
the goodness of fit in terms of $\chi_1^2$,
the error in the extinction fit (see eq.\,\ref{eq:chi1}),
and $\chi_2^2$, the extent to which the abundances
of C, Si, Mg, and Fe required to be depleted in dust
exceed what are available in the ISM
(see eq.\,\ref{eq:chi2}).

We use the Levenberg-Marquardt method
(Press et al.\ 1992) to minimize
$\chi^2=\chi_1^2+\chi_2^2$,
where $\chi_1^2$ gives the error
in the extinction fit:
\begin{equation}\label{eq:chi1}
\chi_1^2 = \sum_i \frac{\left\{
\ln A_{\rm obs}(\lambda_i) 
- \ln A_{\rm mod}(\lambda_i) \right\}^2}
{\sigma_i^2}~~~,
\end{equation}
where $A_{\rm obs}(\lambda_i)$ 
is the observed extinction at wavelength $\lambda_i$,
$A_{\rm mod}(\lambda_i)$ 
is the extinction computed
for the model at wavelength $\lambda_i$
(see eq.\,\ref{eq:extcurv}), 
and $\sigma_i$ characterizes
the weight at  $\lambda_i$.
Following WD01,
we take the weights $\sigma_i^{-1} = 1$ 
for $1.1 < \lambda^{-1} < 8\mum^{-1}$ 
and $\sigma_i^{-1} = 1/3$ for 
$\lambda^{-1} < 1.1\mum^{-1}$.

The term $\chi_2^2$ is a ``penalty''
which keeps the total volumes in the silicate
($\Vs$) and carbonaceous ($\Vg$) grain populations
from grossly violating the interstellar abundance constraints. 
Following WD01, $\chi_2^2$ is measured as
\begin{equation}\label{eq:chi2} 
\chi_2^2 = 0.4 \left\{\max\left(
\Vs/\Vsmax,1\right) -1\right\}^{1.5}
+ 
0.4 \left\{\max\left(
\Vg/\Vgmax,1\right) -1\right\}^{1.5} ~~,
\end{equation}
where $\Vsmax$ and $\Vgmax$ are
the maximum allowable volumes of silicate and
carbonaceous grains imposed by the interstellar
abundance constraints, respectively.

By assuming an olivine-type silicate composition
with a stoichiometry of
Mg$_{2x}$Fe$_{2(1-x)}$SiO$_4$
where $0\simlt x\simlt 1$
(i.e., each Si atom corresponds to four O atoms),
we determine $\Vsmax$ as follows:
\begin{equation}
\Vsmax/\rmH=\frac{
m_{\rm H} \left\{\sidust\mu_{\rm Si}
+\mgdust\mu_{\rm Mg}+\fedust\mu_{\rm Fe}
+ 4 \sidust\mu_{\rm O}\right\}}
{\rho_{\rm sil}} ~~,
\end{equation}
where $\rho_{\rm sil}\approx3.5\g\cm^{-3}$
is the mass density of silicate dust,
$m_{\rm H}\approx 1.67\times 10^{-24}\g$
is the mass of a hydrogen atom,
$\mu_{\rm Si}\approx28$, $\mu_{\rm Mg}\approx24$,
$\mu_{\rm Fe}\approx56$, and $\mu_{\rm O}\approx16$
are the atomic weights of Si, Mg, Fe, and O atoms,
respectively, and $\sidust$, $\mgdust$ and $\fedust$
are the abundances of Si, Mg, and Fe 
(relative to H) locked up in silicate grains, respectively.
By taking the proto-Sun\,+\,GCE abundances
of Si, Mg and Fe (see Table~\ref{tab:abund}),
and assuming that all Si, Mg, and Fe elements
are depleted in silicate grains,
i.e., $\sidust\approx\siism$, 
$\mgdust\approx\mgism$, and
$\fedust\approx\feism$, we obtain 
$\Vsmax\approx4.26\times10^{-27}\cm^3\,\rmH^{-1}$.

Similarly, we determine $\Vgmax$ for carbonaceous grains
as follows:
\begin{equation}
\Vgmax/\rmH = \frac{m_{\rm H}
\left\{\cism-\cgas\right\}\mu_{\rm C}}
{\rho_{\rm C}} ~~,
\end{equation}
where $\mu_{\rm C}\approx12$ is the molecular weight
of carbon grains (on a per C atom basis),\footnote{%
As graphite is a material composed primarily
of carbon atoms (arranged in a hexagonal lattice
structure), this is an accurate approximation.
For PAHs, their hydrogen contents do not add
much to $\mu_{\rm C}$.
As discussed in Li \& Draine (2001),
for compact, pericondensed PAHs,
the hydrogen to carbon ratio
ranges from $\simali$0.5
(for small molecules of fewer than 25 C atoms) 
to $\simali$0.25
(for large PAHs of more than 100 C atoms),
and thus the molecular weight
(per C atom) ranges from
$\mu_{\rm C}\approx12.5$
to $\simali$12.25.
As the bulk mass of the carbonaceous component
is in graphite, we adopt $\mu_{\rm C}\approx12$.
After all, the molecular weight
(per C atom) of PAHs is close to this value. 
}
and $\rho_{\rm C}=2.24\,{\rm g\,cm^{-3}}$
is the density of graphitic materials.
By taking the proto-Sun\,+\,GCE abundance
of C (see Table~\ref{tab:abund}),
and assuming an interstellar gas-phase abundance
of $\cgas\approx198\ppm$ \citep{Hensley2021},
we obtain
$\Vgmax\approx2.30\times10^{-27}\cm^3\,\rmH^{-1}$.

We show in Figure~\ref{fig:bestfit} the best-fit
in the context of adopting the proto-Sun\,+\,GCE
abundances as the interstellar abundances of
dust-forming elements (see Table~\ref{tab:abund}).
The derived size distributions for silicates
and graphite/PAHs are shown in Figure~\ref{fig:dnda}.
We tabulate the best-fit model parameters
in Table~\ref{tab:modpara}.
We note that,
as illustrated in Figure~\ref{fig:dnda},
the model parameters and size distributions
derived here somewhat differ from that of WD01
since we adopt a different set of interstellar abundances
for the dust-forming elements.
As indicated in eq.\,\ref{eq:chi2}, the abundance
difference would affect the goodness of fit
and leads to different model parameters.

Once the grain size distributions are determined
through fitting the interstellar extinction curve,
we can calculate the X-ray scattering and absorption
by dust from eqs.\,\ref{eq:xray_dust_abs},\ref{eq:xray_dust_sca}.
The results are elaborated in \S\ref{sec:results}.

The interstellar extinction modeling also allows
us to infer $\cdust$, the amount of C tied up
in carbonaceous grains (i.e., graphite and PAHs):
\begin{equation}\label{eq:C2H}
\cdust\approx
\frac{\rho_{\rm C} \int_{a_{\rm min}}^{a_{\rm max}} 
\left(4\pi/3\right) a^{3} \left(dn/da\right)_{\rm g}\,da}
{\mu_{\rm C}\,m_{\rm H}} ~~,
\end{equation}
and $\sidust$, $\mgdust$, $\fedust$ as well as
$\odust$, the amounts of Si, Mg, Fe and O depleted
in amorphous silicates:
\begin{equation}\label{eq:Si2H}
\sidust\approx\frac{\rho_{\rm sil} \int_{a_{\rm min}}^{a_{\rm max}} 
\left(4\pi/3\right) a^{3} \left(dn/da\right)_{\rm s}\,da}
{\mu_{\rm sil}\,m_{\rm H}} ~~,
\end{equation}
where $\mu_{\rm sil}$ is the molecular weight of
silicate. By assuming a stoichiometric composition
of MgFeSiO$_4$, we adopt $\mu_{\rm sil}\approx172$
and $\mgdust\approx\sidust$,
$\fedust\approx\sidust$, and
$\odust\approx4\sidust$.
In the context of the proto-Sun\,+\,GCE
abundances for Si, Mg, and Fe,
we indeed have
$\sidust\approx\mgdust\approx\fedust$,
if these elements are essentially completely
depleted in dust.
We tabulate in Table~\ref{tab:modpara}
the derived volumes for silicate grains ($\Vs$)
and carbonaceous grains ($\Vg$).
Table~\ref{tab:dust_abund_Beta_X} lists
the elemental abundances of C, O, Mg, Si, and Fe
required to be locked up in dust, as determined
from the interstellar extinction modeling.
We note that the derived $\mgdust$, $\sidust$,
and $\fedust$ abundances are slightly higher
than the proto-Sun\,+\,GCE reference abundances
listed in Table~\ref{tab:abund}, but still  within
the quoted uncertainties. 

Finally, we derive the ``depletion factor''
$\left(1-\beta_{\rm X}\right)$,
where X\,=\,C, Si, Mg, Fe, and O,
from $\cdust$, $\sidust$, $\mgdust$, $\fedust$
and $\odust$, respectively.
The ``depletion factors'' are crucial for
calculating the X-ray absorption by gas
(see eq.\,\ref{eq:xray_gas_abs_1}).
The X-ray absorption arising from gaseous
species will be discussed in \S\ref{sec:gas}
and the results will be presented in \S\ref{sec:results}.

\begin{table}
\centering
\caption{\label{tab:modpara}
Model Parameters for Silicate and Graphite Grains}
\begin{tabular}{lr}
\hline\hline
Parameter & Value \\
\hline
$\alpha_{\rm g}$ & $-1.77$ \\
$\beta_{\rm g}$ & $-0.129$ \\
$a_{\rm t,g}$ ($\mu$m) & $0.0121$ \\
$a_{\rm c,g}$ ($\mu$m) & $1.074$ \\
$C_{\rm g}$  & $6.41\times10^{-12}$ \\
$\alpha_{\rm s}$ & $-2.31$ \\
$\beta_{\rm s}$ & $0.509$ \\
$a_{\rm t,s}$ ($\mu$m) & $0.213$ \\
$a_{\rm c,s}$ ($\mu$m) & $0.100$ \\
$C_{\rm s}$ & $4.56\times 10^{-14}$ \\
$\Vgmax/\rmH$ (${\rm cm}^3\,\rmH^{-1}$) & $2.303\times10^{-27}$ \\
$\Vsmax/\rmH$ (${\rm cm}^3\,\rmH^{-1}$) & $4.255\times10^{-27}$ \\
$\Vg/\rmH$ (${\rm cm}^3\,\rmH^{-1}$) & $2.196\times10^{-27}$ \\
$\Vs/\rmH$ (${\rm cm}^3\,\rmH^{-1}$) & $4.378\times10^{-27}$ \\
$\chi^{2}_{1}$ & $0.606$ \\
$\chi^{2}_{2}$ & $0.002$ \\
\hline
\end{tabular}
\end{table}

\section{Gaseous Species as an X-ray Absorber}\label{sec:gas}
For the gaseous species, their photoionization cross
sections are dependent on their ionization stages.
Large parts of the ISM are moderately ionized,
and in principle, we should account for
the ionized phase of the ISM.  
While Wilms et al.\ (2000) only considered
the neutral phase of the ISM, as elaborated below,
we will also consider carbon ions as well.

The ionization states of the elements in the Galactic
diffuse ISM or HI regions are determined primarily
by their first ionization potentials
(e.g., see Jiang et al.\ 2026, Zhang et al.\ 2026).
For carbon atoms, with a first ionization potential
of 11.26$\eV$, they are expected to be singly ionized
(C\,II) in HI regions. In the Milky Way galaxy,
as mentioned in Footnote~\ref{ftnt:fracH2},
$\simali$90\% of the hydrogen nucleons are
in atomic form and $\simali$10\% in molecular form.
In regions where hydrogen is mostly molecular,
gaseous carbon will be atomic (CI) and/or molecular
(e.g., CO). 
By approximating the photoelectric absorption
cross sections of CI for that of the C atoms in CO,
we adopt a number fractions of
90\% for C\,II and 10\% for C\,I.
For gaseous He, N, and O, with an ionization potential
of 24.59, 14.53, and 13.62$\eV$, respectively,
they will be mostly atomic in HI regions.
Therefore, we will only consider the photoionization
cross sections of atomic He, N, and O.
For Mg, Si, and Fe, if not depleted in solids,
they will be singly ionized in HI regions
as their ionization potentials are only
7.65, 8.15 and 7.90$\eV$, respectively.
However, as shown in Table~\ref{tab:dust_abund_Beta_X},
our interstellar extinction model requires
all Mg, Si, and Fe elements to be completely
depleted in silicate grains.
Therefore, we do not need to consider their ions.

For many other metal elements
(e.g., Al, Ca, Ti, Cr, Mn, Co, Ni),
it is well recognized that they are also
severely depleted in dust, i.e.,
$\left(1-\beta_{\rm X}\right)\approx 0$.
However, we have little information regarding
the compositions and size distributions
of the dust grains in which these elements reside.
We therefore assume them to be all in gas phase,
i.e., $\left(1-\beta_{\rm X}\right)\approx 1.0$.

Nevertheless, it is well established that
the effective X-ray absorption of solids,
on a per unit atom basis,
is reduced due to self-shielding within the grain.
Treating Al, Ca, Ti, Cr, Mn, Co, Ni and other metal
elements as gas atoms will somewhat overestimate
their X-ray absorption.
However, as these elements are much less
abundant than H, He, C, N, O, Si, Mg, and Fe,
their contribution to the interstellar X-ray absorption
is rather small. Indeed, if we neglect entirely those
elements with $\xism < 10^{-5}$, the resulting X-ray
absorption is only reduced by $\simali$5\%.
Therefore, our simplified treatment of Al, Ca, Ti, Cr,
Mn, Co, Ni, and other metals will not affect our results.

To summarize, we compute the X-ray absorption resulting
from gaseous species as follows:
\begin{eqnarray}\label{eq:xray_gas_abs2}
\nonumber
\sigmagas/\rmH &=& \sum_{{\rm X}}
\xism\left(1-\beta_{\rm X}\right)
                   {\bigsigma_{\rm bf}({\rm X})}\\
\nonumber
&&+ \cism \left(1-\beta_{\rm C}\right)
\left\{f({\rm CI}) \bigsigma_{\rm bf}({\rm CI})
+f({\rm CII}) \bigsigma_{\rm bf}({\rm CII})\right\}\\
&&+ f({\rmHI})\bigsigma_{\rm bf}(\rmHI)
\,+\,f(\rmHH)\bigsigma_{\rm bf}(\rmHH) ~~,
\end{eqnarray}
where the sum is over X\,=\,He, O, N, ...,
$f({\rm CI})\approx0.1$ and $f({\rm CII})\approx0.9$
are the relative amounts of carbon atoms
and ions with respect to the total carbon nucleon, respectively,
and $\bigsigma_{\rm bf}({\rm CI})$
and $\bigsigma_{\rm bf}({\rm CII})$ 
are the photonization cross sections
of carbon atoms and ions, respectively.
In computing the X-ray absorption caused by
gaseous species, we take the photoionization
cross sections of Band et al.\ (1990) for atomic hydrogen,
of Yan et al.\ (1998) for He and H$_2$,
of Gatuzz et al.\ (2015) for O, N, and Ne,
and of Verner et al.\ (1993) for all other elements.

\section{Results and Discussion}\label{sec:results}
With the grain size distributions determined
from fitting the interstellar extinction curve
(see \S\ref{sec:dust} and Figure~\ref{fig:dnda}),
we now calculate the absorption and scattering
cross sections of amorphous silicates and graphite/PAHs
in the soft X-ray energy range of 0.1--10\,keV,
utilizing Mie theory, based on the technique
developed by Wang \& van de Hulst (1991) 
which is accurate for $2\pi a/\lambda$
up to 5$\times$10$^5$. Again, we adopt the dielectric
functions of astronomical silicates and graphite
from \citet{Draine2007}. 
For PAHs, their optical
properties at $\lambda \simlt 0.06\mum$ were
set to be the same as that of graphite
on a per C atom basis (see eq.\,5 of Li \& Draine 2001).
The total X-ray absorption and scattering per H nucleon
caused by dust are obtained by integrating the absorption
and scattering cross sections over the size distributions
(see eqs.\,\ref{eq:xray_dust_abs},\ref{eq:xray_dust_sca}),
while the X-ray extinction is the sum of the absorption
and scattering (see eq.\,\ref{eq:xray_dust_ext}).

\begin{figure*}[htp]
\vspace{3mm}
\begin{center}
\includegraphics[width=12cm,angle=0]{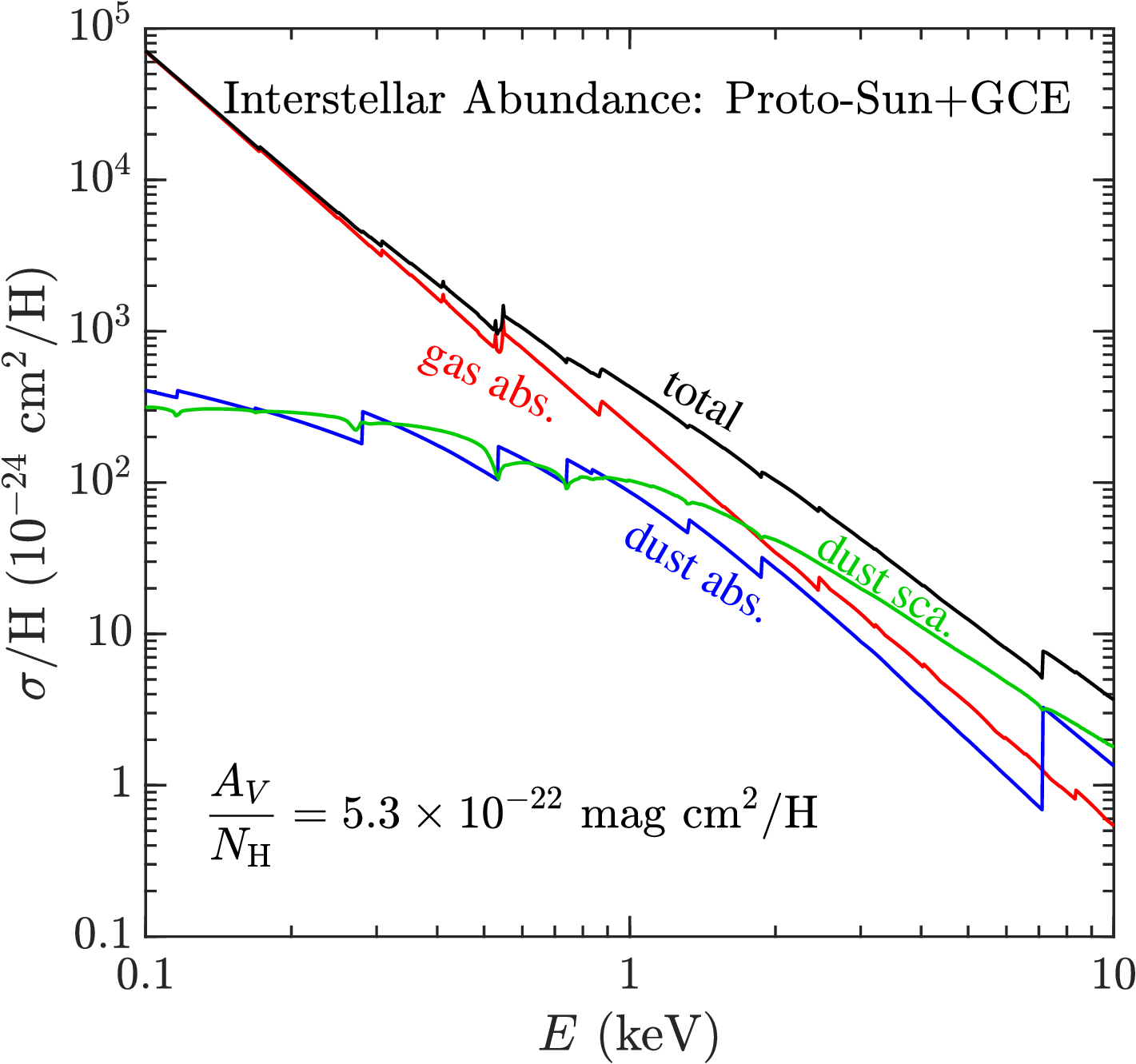}
\end{center}
\vspace{-4mm}
\caption{\label{fig:xray_dust+gas}
Interstellar X-ray extinction per H nucleon
(black line) as a combination of
gas absorption $\sigmagas$ (red line), 
dust absorption $\sigmadustabs$ (blue line),
and dust scattering $\sigmadustsca$ (green line).
}
\end{figure*}

Figure~\ref{fig:xray_dust+gas} shows
the X-ray absorption and scattering
per H nucleon arising from the silicate
and graphite/PAHs mixture.
At low energies ($E \lesssim 0.3\keV$),
the extinction is dominated by gas absorption, 
while the contribution from dust becomes
appreciable at $E \gtrsim 0.3\keV$. 
Between $0.3\keV \lesssim E \lesssim 1\keV$,
dust scattering is comparable to dust absorption,
and at higher energies ($E \gtrsim 1\keV$),
dust extinction dominates over gas absorption.
This demonstrates the important role of dust
in attenuating X-ray photons.

\begin{table}[h!]
\centering
\caption{Elemental Abundances Locked up
               in Dust and the Depletion Factors}
\label{tab:dust_abund_Beta_X}
\begin{tabular}{lccccr}
\hline\hline
Element & (X/H)$_{\rm dust}$
& $\left(1-\beta_X\right)$ \\
&  (ppm) &   \\
\hline
C    & 246     & 0.46   \\
O    & 215     & 0.72   \\
Mg  & 53.7    &  0      \\
Si   & 53.7    &  0  \\
Fe  & 53.7    &  0  \\
\hline
\end{tabular}
\end{table}

As elaborated in \S\ref{sec:dust},
the interstellar extinction modeling
also provides constraints on
the ``depletion factor''
$\left(1-\beta_{\rm X}\right)$
which is required for calculating
the X-ray absorption of gas
(see eq.\,\ref{eq:xray_gas_abs2}).
We deduce $\cdust$,
the amounts of C tied up in carbonaceous grains
(i.e., graphite and PAHs),
from eq.\,\ref{eq:C2H},
and $\sidust$, $\mgdust$,
$\fedust$ and $\odust$,
the amounts of Si, Mg, Fe and O
locked up in amorphous silicate grains,
from eq.\,\ref{eq:Si2H}.
With $\cdust$, $\sidust$, $\mgdust$,
$\fedust$ and $\odust$ determined,
we derive $\left(1-\beta_{\rm X}\right)$
and tabulate them in Table~\ref{tab:dust_abund_Beta_X},
where X\,=\,C, Si, Mg, Fe, and O.
For all other elements, we assume no
depletion in dust, i.e.,  $\beta_{\rm X}=0$.
Undoubtedly, as already mentioned in \S\ref{sec:gas},
this is a simplified assumption as it is well
recognized that many metal elements
such as Al, Ca, and Ni are also severely depleted
in dust (e.g., see Jenkins 2009).
As we have little information
about the exact composition of the solids
in which these elements reside,
we simply calculate their X-ray absorption
by assuming that they are all in gas phase.
Nevertheless, as discussed in \S\ref{sec:gas},
this does not appreciably affect our results
since their abundances are much lower than
those of the major dust-forming elements
(i.e., C, Si, Mg, Fe, and O).

With the  ``depletion factors''
$\left(1-\beta_{\rm X}\right)$ determined,
we now calculate the X-ray absorption
arising from gaseous species as elaborated
in \S\ref{sec:gas}. The resulting X-ray absorption
by gas is also shown in Figure~\ref{fig:xray_dust+gas}. 
Apparently, at low energies (say, $E\simlt0.3\keV$),
the interstellar X-ray absorption is dominated
by gas and the contribution from dust only
becomes appreciable at $E\gtsim0.3\keV$.
At $E\gtsim0.8\keV$, the contribution to
the interstellar X-ray extinction by dust
becomes comparable to that of gas.
At higher energies (say, $E\gtsim1\keV$),
dust extinction becomes more important than gas absorption.
This clearly demonstrates the important
role of dust in attenuating X-ray photons.

\begin{figure*}[htp]
\vspace{3mm}
\begin{center}
\includegraphics[width=11.6cm,angle=0]{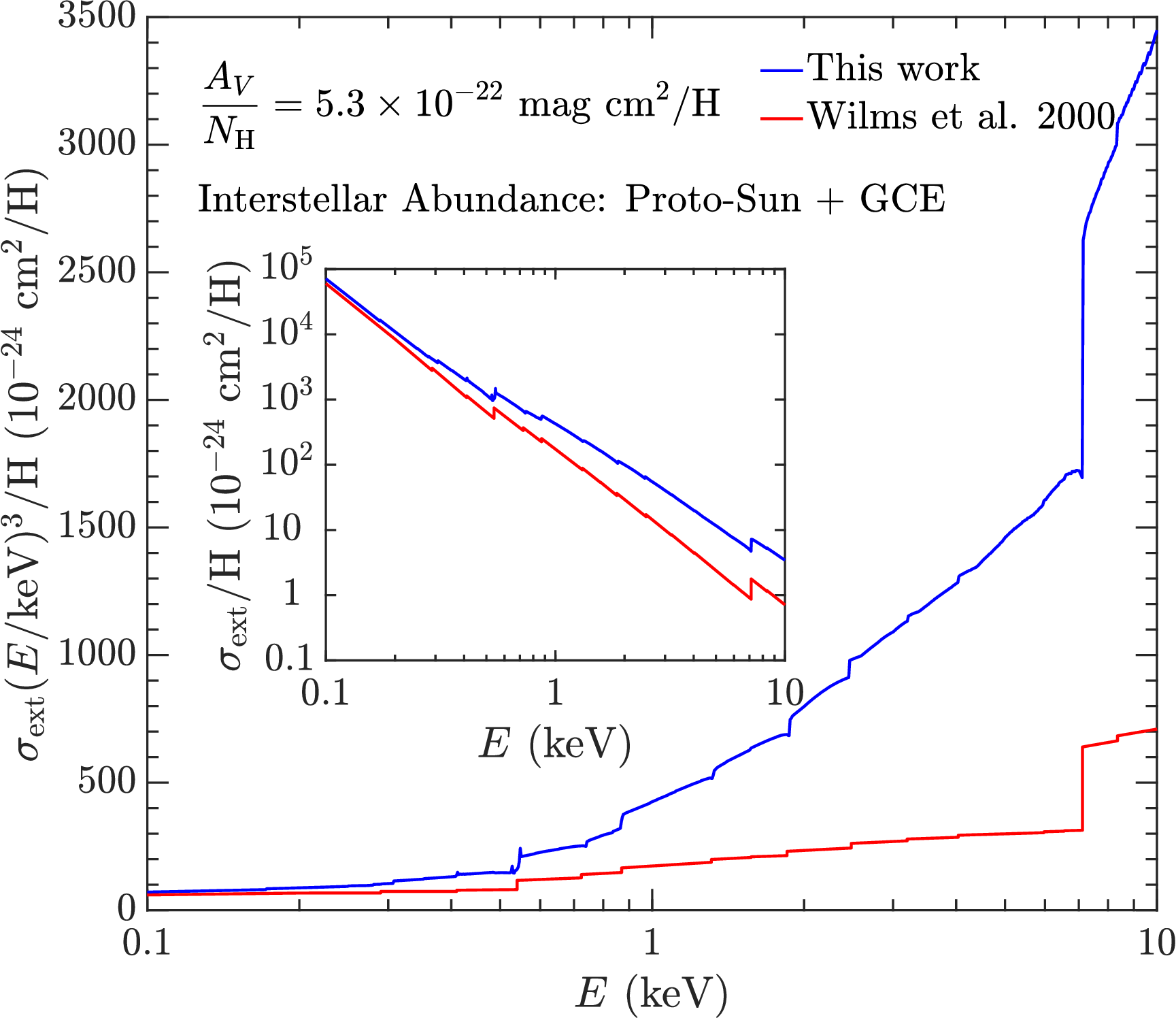}
\end{center}
\vspace{-4mm}
\caption{\label{fig:yll_vs_wilms}
Comparison of the interstellar X-ray extinction
calculated here with that of Wilms et al.\ (2000).  
For clarity, the X-ray extinction has been
multiplied by $\left(E/{\rm keV}\right)^3$.
The inset shows a comparison between
the unmultiplied quantities.
Clearly, the interstellar X-ray extinction
derived here significantly exceeds that
of Wilms et al.\ (2000) at $E\gtsim0.5\keV$.
         }
\end{figure*}

We compare in Figure~\ref{fig:yll_vs_wilms}
the interstellar X-ray extinction calculated here
with that of Wilms et al.\ (2000).  
For clarity, the X-ray extinction has been
multiplied by $\left(E/{\rm keV}\right)^3$.
It is clear that the interstellar X-ray extinction
derived here significantly exceeds that
of Wilms et al.\ (2000) at $E\gtsim0.5\keV$,
and this effect becomes increasingly more
prominent toward higher energies.
More specifically, the X-ray extinction derived here
exceeds that of Wilms et al.\ (2000)
by a factor of $\simali$1.7 at 0.5$\keV$,
$\simali$2.1 at 1$\keV$, 
$\simali$3.2 at 2$\keV$,
$\simali$4.0 at 5$\keV$,
$\simali$5.5 at 7$\keV$, and
$\simali$4.9 at 10$\keV$.

As mentioned in \S\ref{sec:intro}, these increases
can be qualitatively understood in terms of
the interstellar abundances and dust scattering.
For example, the Mg, Si and Fe abundances 
of Wilms et al.\ (2000) are lower than that of
the proto-Sun\,+\,GCE reference standard
by a factor of $\simali$2. This implies that
Wilms et al.\ (2000) would have underestimated
the silicate dust quantity
(and therefore $\sigmadust/\rmH$)
by a factor of $\simali$2.
On the other hand, the neglect of dust scattering
further underestimates $\sigmadust/\rmH$.

The importance of accounting for
the scattering of X-rays by dust has previously
been investigated by \citet{Draine2003},
\citet{Corrales2016} and \citet{HoffmanDraine2016}.
\citet{Corrales2016} computed the interstellar
X-ray absorption and scattering from a mixture
of silicate and graphite grains, assuming a mass
mixing ratio of 60\% silicate to 40\% graphite
and a MRN-type power-law size distribution of
$dn/da\propto a^{-3.5}$
over $50\Angstrom <a<0.25\mum$
(Mathis et al.\ 1977) for both dust components.
Similar to \citet{Corrales2016},
\citet{Draine2003} and \citet{HoffmanDraine2016}
also computed the interstellar X-ray absorption
and scattering of silicate and graphite grains,
but employing the WD01 size distributions.
Compared to these earlier efforts,
in this work, we made use of
the latest results on the interstellar abundances
and derived the depletion factor for
each element, after accounting for
the elemental depletions in dust
determined from modeling the near-IR
to far-UV extinction curve.

It is worth noting that, X-ray scattering by dust
is strongly forward-peaked and occurs only
over a very small, arcmin-scale angle. 
Therefore, the effective dust scattering
depends on both the dust geometry
and the size of the region used to extract
the data of an X-ray point source
(Smith et al.\ 2016).
As Corrales et al.\ (2016) pointed out,
practically, the need to include X-ray scattering
by dust in the X-ray extinction model is situational,
depending on the imaging resolutions of
the X-ray observing facilities,
as well as the geometric effects of the scattering dust.
In other words, whether scattering contributes to
X-ray attenuation depends  on the relation
between the X-ray scattering angular scale
and the extraction region: 
X-ray photons scattered outside the extraction region
are effectively removed from the line of sight,
while those within it may be partially recovered.
The X-ray extinction model presented here
does not accounted for the angular redistribution
of X-ray photons and their partial recovery
within the source aperture.
Therefore, it represents the limiting case
in which scattered photons are all removed 
from the direct beam. 
In practical observations, partial recovery of
scattered photons can reduce the effective
extinction, depending on the observational configuration.

We also note that \citet{Moutard2026} recently
determined the interstellar Fe abundance
by fitting the Fe-L absorption edges seen
in the high-resolution X-ray absorption
spectra of background X-ray sources.
While the derived abundance of
$\feism\approx30.3\ppm$
is close to the latest solar Fe/H
abundance reported by Asplund et al.\ (2021)
and Lodders et al.\ (2025),
it is significantly lower than
the proto-Sun\,+\,GCE abundance
of $\feism\approx52.5\ppm$
adopted here (see Table~\ref{tab:abund}).
However, one should keep in mind that
the X-ray measurements of Moutard et al.\ (2026)
were a direct measurement of the column density
of Fe atoms ($N_{\rm Fe}$), both in dust and in gas.
To convert $N_{\rm Fe}$ into Fe/H,
Moutard et al.\ (2026) used $\NH$---the hydrogen
column density---derived from X-ray modeling,
which relies on assumptions about the relative
abundance mixture of the other elements.
Indeed, by adopting the underestimated
X-ray absorption data of Wilms et al.\ (2000),
Moutard et al.\ (2026) would have overestimated
$\NH$ and therefore underestimated Fe/H.
Therefore, a more robust quantity for comparison
is the iron to neon ratio (Fe/Ne)
since Moutard et al.\ (2026) also measured
the column density of Ne atoms
($N_{\rm Ne}$) through the Ne-K absorption edges.
As shown in Table~\ref{tab:Fe_Ne_ratio},
the Fe/Ne ratio observationally
derived by Moutard et al.\ (2026)
is actually close to that of
the proto-Sun\,+\,GCE abundance standard,
thereby supporting the suitability of
the proto-Sun\,+\,GCE abundances as
the interstellar reference abundances.

\begin{table}[htbp]
{\footnotesize\centering
\caption{Fe and Ne Abundances
(Relative to $10^6$ H Atoms) from Various
Reference Standards (See Table~\ref{tab:abund})
and Derived from the Fe-L and Ne-K Absorption
Edges (Moutard et al.\ 2026)}
\label{tab:Fe_Ne_ratio}
\begin{tabular}{lccccr}
\hline
Element & Sun$^{a}$ & Sun$^{b}$ & Proto-Sun$^{b}$
& Proto-Sun\,+\,CGE$^{c}$ & Moutard et al.\ (2026)$^{d}$ \\
\hline
 Fe  & $28.8 \pm 2.7$ & ${30.9 \pm 0.7}$
& ${38.0 \pm 0.9}$ & ${52.5 \pm 1.2}$
& ${30.3 \pm 1.1}$\\
Ne & ${114.8 \pm 13.2}$ & ${141.3 \pm 39}$
& ${173.8 \pm 48}$ & ${190.5 \pm 52.6}$ 
& ${102.8 \pm 2.8}$\\
\hline
Fe/Ne & ${0.251 \pm 0.037}$
& ${0.219 \pm 0.061}$
& ${0.219 \pm 0.061}$
& ${0.276 \pm 0.076}$
& ${0.295 \pm 0.013}$\\
\hline
\end{tabular}
\parbox{0.95\linewidth}{
(a) Asplund et al.\ (2021);
(b) Lodders et al.\ (2025);
(c) Lodders et al. (2025), Chiappini et al. (2003);
(d) Moutard et al.\ (2026) derived the Fe and Ne
abundances from the high-resolution X-ray absorption
spectra of background X-ray sources.
}
}
\end{table}

Finally, the interstellar X-ray extinction (absorption
and scattering) model developed in this work
is publicly available on {\tt GitHub}.\footnote{%
  \texttt{
    https://github.com/yanll-xray/xray-extinction-model.
  }
  }
It can be incorporated into modern X-ray spectral
fitting softwares such as {\tt XSPEC}
as a multiplicative table model. 
The package includes model files for
the X-ray absorption, scattering
and extinction cross sections per H nucleon
as a function of energy, as well as example scripts
demonstrating how to apply our model
to correct the X-ray observations for
Galactic interstellar X-ray extinction.

\section{Summary}\label{sec:summary}
We have updated the interstellar X-ray absorption
and scattering by making use of updated atomic
cross sections, and taking into account 
realistic interstellar elemental abundances
and the scattering of X-rays by dust grains.
It is found that the X-ray absorption and scattering
derived here are considerably higher than those
commonly used by the X-ray astronomical community.

\acknowledgments{%
We thank B.T.~Draine and A.N.~Witt
for helpful discussions.
We also thank the anonymous referee
for his/her insightful and constructive
comments and suggestions that have
significantly improved the quality and
presentation of this work.
LLY is supported in part by
the National Natural Science Foundation
of China (No.\,12473022),
CMS-CSST-2021-A09,
and the Talent Programs of
the Anhui Provincial Department of Education
(Nos.\,YQZD2023053, 2024AH030011).
}


\end{document}